\documentclass[aps,prd,onecolumn,superscriptaddress,showpacs]{revtex4}

\usepackage{graphicx}
\usepackage{dcolumn}
\usepackage{bm}
\usepackage{color}
\usepackage{amsmath}
\usepackage{amstext}
\usepackage{pstricks}
\usepackage{color}
\newcommand{\gsim}{ \mathop{}_{\textstyle \sim}^{\textstyle >} }
\newcommand{\lsim}{ \mathop{}_{\textstyle \sim}^{\textstyle <}}
\newcommand{\vev}[1]{ \left\langle {#1} \right\rangle }
\newcommand{\sigmav}{\langle\sigma v\rangle}

\newcommand{\be}{\begin{eqnarray}}
\newcommand{\ee}{\end{eqnarray}}
\newcommand{\tev}{\rm \, TeV}
\newcommand{\gev}{\rm \, GeV}
\newcommand{\mev}{\rm \, MeV}
\newcommand{\kev}{\rm \, keV}
\newcommand{\kms}{\ensuremath {\rm km/s}}
\newcommand{\cm}{\rm \, cm}

\begin{document}

\title{A Theory of Dark Matter}
\author{Nima Arkani-Hamed}
\affiliation{School of Natural Sciences, Institute for Advanced Study, Princeton, NJ 08540, USA}

\author{Douglas P. Finkbeiner}
\affiliation{Harvard-Smithsonian Center for Astrophysics, 60 Garden St., Cambridge, MA 02138, USA}

\author{Tracy R. Slatyer}
\affiliation{Physics Department, Harvard University, Cambridge, MA 02138, USA}

\author{Neal Weiner}
\affiliation{Center for Cosmology and Particle Physics, Department of Physics, New York University,
New York, NY 10003, USA}

\date{\today}

\begin{abstract}
We propose a comprehensive theory of dark matter that explains the
recent proliferation of unexpected observations in high-energy
astrophysics. Cosmic ray spectra from ATIC and PAMELA require a WIMP
with mass $M_\chi \sim 500 - 800\gev$ that annihilates into leptons
at a level well above that expected from a thermal relic. Signals
from WMAP and EGRET reinforce this interpretation.  Limits on $\bar
p$ and $\pi^0$-$\gamma$'s constrain the hadronic channels allowed
for dark matter.  Taken together, we argue these facts imply the
presence of a new force in the dark sector, with a Compton
wavelength $m_\phi^{-1} \gsim 1 \gev^{-1}$. The long range allows a
Sommerfeld enhancement to boost the annihilation cross section as
required, without altering the weak-scale annihilation cross section
during dark matter freeze-out in the early universe. If the dark
matter annihilates into the new force carrier $\phi$, its low mass
can make hadronic modes kinematically inaccessible, forcing decays
dominantly into leptons. If the force carrier is a non-Abelian gauge
boson, the dark matter is part of a multiplet of states, and
splittings between these states are naturally generated with size
$\alpha m_\phi \sim$ MeV, leading to the eXciting dark matter (XDM)
scenario previously proposed to explain the positron annihilation in
the galactic center observed by the INTEGRAL satellite;  the light
boson invoked by XDM to mediate a large inelastic scattering cross
section is identified with the $\phi$ here. Somewhat smaller
splittings would also be expected, providing a natural source for
the parameters of the inelastic dark matter (iDM) explanation for
the DAMA annual modulation signal. Since the Sommerfeld enhancement
is most significant at low velocities, early dark matter halos at
redshift $\sim10$ potentially produce observable effects on the
ionization history of the universe. Because of the enhanced cross
section, detection of substructure is more probable than with a
conventional WIMP. Moreover, the low velocity dispersion of dwarf
galaxies and Milky Way subhalos can increase the substructure
annihilation signal by an additional order of magnitude or
more.\end{abstract}


\maketitle


\section{PAMELA/ATIC and New Dark Forces}
Thermal WIMPs remain one of the most attractive candidates for dark
matter. In addition to appearing generically in theories of
weak-scale physics beyond the standard model, they naturally give
the appropriate relic abundance. Such particles also are very
promising in terms of direct and indirect detection, because they
must have some connection to standard model particles.

Indirect detection is particularly attractive in this respect. If dark matter annihilates to some set of standard model states, cosmic ray detectors such as PAMELA, ATIC and Fermi/GLAST have the prospect of detecting it. This is appealing, because it directly ties the observable to the processes that determine the relic abundance.

For a weak-scale thermal particle, the relic abundance in the case of $s$-wave annihilation is approximately set by
\be
\Omega h^2 \simeq 0.1 \times \left( \frac{\sigmav_{\rm freeze}}{3 \times 10^{-26} {\rm\,cm^{3} s^{-1}}}\right)^{-1}.
\ee
For perturbative annihilations, $s$-wave dominates in the late universe, so this provides an approximate upper limit on the signal that can be observed in the present day. Such a low cross section makes indirect detection, whereby the annihilation products of dark matter are detected in cosmic ray detectors, a daunting task.

However, recent experiments have confirmed the longstanding suspicion
that there are more positrons and electrons at 10s-100s of GeV than can be explained by supernova shocks and interactions of cosmic ray protons with the ISM.  The experiments are
\begin{itemize}
\item{{\bf PAMELA} \\ The Payload for Antimatter Matter Exploration
    and Light-nuclei Astrophysics has reported results
    \cite{Adriani:2008zr} indicating a sharp upturn in
    the positron fraction ($e^+/(e^++e^-)$) from $10-100\gev$, counter
    to what is expected from high-energy cosmic rays interacting with
    the interstellar medium (ISM). This result confirms excesses seen
    in previous experiments, such as HEAT
    \cite{barwick:1997ig,Beatty:2004cy} and AMS-01
    \cite{Aguilar:2007yf}. One possible explanation for this is dark
    matter annihilation into $e^+e^-$ \cite{Kane:2002nm,
      Hooper:2003ad, Cholis:2008qq}, but this requires a large cross section \cite{Baltz:2001ir}.} 
\item{{\bf ATIC} \\ The Advanced Thin Ionization Calorimeter is a balloon-borne cosmic ray detector which studies electrons and positrons (as well as other cosmic rays) up to $\sim\tev$ energies, but cannot distinguish positrons and electrons. The primary astrophysical sources of high-energy electrons are expected to be supernovae: electrons are accelerated to relativistic speeds in supernova remnants and then diffuse outward. The ATIC-2 experiment reported a $4-6 \sigma$ excess (over a simple power law) in its $e^++e^-$ data \cite{Chang:2008zz} at energies of $\sim 300 -800 \gev$, with a sharp cutoff in the $600-800 \gev$ range. Dark matter would seem a natural candidate for this as well, with its mass scale determining the cutoff.}
\item{{\bf WMAP} \\ Studies of the WMAP microwave emission from the galactic center show a hard component not spatially correlated with any known galactic emission mechanism. This ``WMAP haze'' \cite{Finkbeiner:2003im, Dobler:2007wv} can be explained as synchrotron radiation from electrons and positrons produced from dark matter annihilation in the galactic center \cite{Hooper:2007kb}.}
\item{{\bf EGRET} \\
Gamma-ray measurements in the galactic center
(inner $5^\circ$) provide hints of an excess at $10-50\gev$
\cite{strong:2005}. Strong et al.
reanalyzed EGRET data and found a harder spectrum at
these energies than previously derived,
using the improved EGRET sensitivity estimates of
\cite{thompson:2005}.  Despite poor spatial resolution, Strong et al. found an
excess in this energy range above the expected $\pi^0$ gamma-ray emission
from cosmic ray protons interacting with the interstellar medium (see Fig. 8 of
 \cite{strong:2005}). Such $\gamma$'s could naturally arise from
inverse-Compton scattering of high-energy electrons and positrons
off of starlight and the cosmic microwave background (CMB).}
\end{itemize}

Taken together, these make a compelling case for excessive electronic
production in the galaxy. While individual astrophysical explanations
may exist for each signal (pulsar wind nebulae for PAMELA and ATIC \cite{1995A&A...294L..41A,Hooper:2008kg,Yuksel:2008rf},
for instance, supernovae for the WMAP haze \footnote{Although, it
should be noted that such an explanation seems to fail because the
spectrum would not be sufficiently hard; see Fig. 6 of \cite{Dobler:2007wv}.}), the data cry out for a
unified explanation. Dark matter annihilations provide an appealing
candidate.

In addition to the above, there are two other anomalies that are worth
mentioning. The INTEGRAL 511 keV signal indicates $\sim
3\times10^{42} e^+/$s annihilating in the galactic center, far more
than expected from supernovae. The spectrum suggests that these
positrons are injected with relatively low energies ($E \lsim {\rm few
\mev}$), and so form a distinct population from those above. eXciting
dark matter (XDM) \cite{Finkbeiner:2007kk}
can naturally explain such low energy
positrons with $\sim 1 \mev$ excited states of the dark matter, while
still producing the high-energy positrons via annihilation
\cite{Cholis:2008vb, Cholis:2008wq}.
Lastly, there is the DAMA/LIBRA indication of an annual
modulation consistent with that expected from dark matter induced
nuclear scattering \cite{Bernabei:2008yi}.
Such a signal is difficult to reconcile with null
results of other experiments, but can be reconciled with $\sim 100
\kev$ excited states in the ``inelastic dark matter'' (iDM) scenario
\cite{Smith:2001hy, Tucker-Smith:2004jv, Chang:2008gd}.
Although we are motivated by the specific signals above (PAMELA/ATIC as well as the haze and EGRET), the picture we are led to for explaining them naturally incorporates the necessary ingredients to explain the INTEGRAL and DAMA signals as well.

Focusing on only the high-energy positrons and electrons, there are a number of challenges to any model of dark matter. These are:
\begin{itemize}
\item{\bf A large cross section} \\ Studies of PAMELA and ATIC signals seem to require a cross section {\em much} larger than what is allowed by thermal relic abundance. Boost factors of $\mathcal{O}(100)$ or more above what would be expected for a thermal WIMP are required to explain these excesses \cite{Cholis:2008hb,Cirelli:2008pk}.
\item{\bf A large cross section {\em into leptons}} \\ Typical annihilations via Z bosons produce very few hard leptons. Annihilations into W bosons produce hard leptons, but many more soft leptons through the hadronic shower. Higgs bosons and heavy quarks produce even softer spectra of leptons, all of which seem to give poor fits to the data. At the same time, absent a leptophilic gauge boson, it is a challenge to construct means by which dark matter would annihilate directly to leptons.
\item{\bf A {\em low} cross section into hadrons}\\ Even if a suitably high annihilation rate into leptons can be achieved, the annihilation rate into hadronic modes must be low. Limits from diffuse galactic gamma rays \cite{Baltz:2008wd}, as well as gamma rays from the center of the galaxy constrain the production of $\pi^0$'s arising from the dark matter annihilation. PAMELA measurements of antiprotons tightly constrain hadronic annihilations as well \cite{Cirelli:2008pk}. Consequently, although quark and gauge boson annihilation channels may occur at some level, the dominant source of leptons must arise through some other channel.
\end{itemize}

The combination of these issues makes the observed high-energy
anomalies -- especially ATIC and PAMELA -- difficult to explain with
thermal dark matter annihilation. However, we shall see that the inclusion of
a new force in the dark sector simultaneously addresses all of these
concerns.

\vskip 0.2in
{\noindent \bf New Forces in the Dark Sector}

\vskip 0.1in
A new interaction for the dark sector can arise naturally in a
variety of theories of physics beyond the standard model, and is
thus well motivated from a theoretical point of view. Although there
are strong limits on the self-interaction scattering cross section
from structure formation \cite{Spergel:2000, Dave:2001}, the presence
of {\em some} new force-carrying boson should be expected, with only
the mass scale in question. A light boson could arise naturally if its
mass scale is generated radiatively \cite{nimaneal}, or if it were a
pseudo-goldstone boson.

One of the important modifications that can arise with a new light boson is an enhancement of the annihilation cross section via a mechanism first described by Sommerfeld \cite{sommerfeld}. The presence of a new force carrier can distort the wave function of the incoming particles away from the plane-wave approximation, yielding significant enhancements (or suppressions) to annihilation cross sections
\footnote{The importance of this effect in the context of multi-TeV scale dark matter interacting via W and Z bosons was first discussed by \cite{2005PhRvD..71f3528H}, and more recently emphasized with regards to PAMELA in the context of ``minimal dark matter,'' with similar masses by \cite{2007NuPhB.787..152C, Cirelli:2008jk}.}.
Equivalently, ladder diagrams involving multiple exchanges of the force carrier must be resummed (Fig. 1).  As we shall describe, the Sommerfeld enhancement can only arise if the gauge boson has a mass $m_\phi \lsim \alpha M_{DM} \sim$ few GeV. Thus, with the mass scale of $\sim 800\gev$ selected by ATIC, and the large cross sections needed by both ATIC and PAMELA, the mass scale for a new force carrier is automatically selected. Interactions involving W and Z bosons are insufficient at this mass scale.

Once this new force carrier $\phi$ is included, the possibility of a new annihilation channel $\chi \chi \rightarrow \phi \phi$ opens up, which can easily be the dominant annihilation channel. Absent couplings to the standard model, some of these particles could naturally be stable for kinematical reasons; even small interactions with the standard model can then lead them to decay only into standard model states. If they decay dominantly into leptons, then a hard spectrum of positrons arises very naturally. Motivated by the setup of XDM, Cholis, Goodenough and Weiner  \cite{Cholis:2008vb} first invoked this mechanism of annihilations into light bosons to provide a simple explanation for the excesses of cosmic ray positrons seen by HEAT without excessive antiprotons or photons. Simple kinematics can forbid a decay into heavier hadronic states, and as we shall see, scalars lighter than $\sim 250 \mev$ and vectors lighter than $\sim \gev$ both provide a mode by which the dark matter can dominantly annihilate into very hard leptons, with few or no $\pi^0$'s or antiprotons.

In the following sections, we shall make these points more concretely and delineate which ranges of parameters most easily explain the data; but the essential point is very simple: if the dark matter is $\mathcal{O}(800\gev)$ and interacts with itself via a force carrier with mass $m_\phi \sim \gev$, annihilation cross sections can be considerably enhanced at present times via a Sommerfeld enhancement, far exceeding the thermal freeze-out cross section.
If that boson has a small mixing with the standard model, its mass scale can make it kinematically incapable of decaying via a hadronic shower, preferring muons, electrons and, in some cases, charged pions, and avoiding constraints from $\pi^0$'s and antiprotons. If the force-carriers are non-Abelian gauge bosons, we shall see that other anomalies may be incorporated naturally in this framework, explaining the INTEGRAL 511 keV line via the mechanism of ``eXciting dark matter'' (XDM), and the DAMA annual modulation signal via the mechanism of ``inelastic dark matter'' (iDM). We shall see that the excited states needed for both of these mechanisms
[$\mathcal{O}(1\mev)$ for XDM and $\mathcal{O}(100\kev)$ for iDM]
arise naturally with the relevant mass splitting generated radiatively at the correct scale.

\section{Sommerfeld enhancements from new forces}
A new force in the dark sector can give rise to the large annihilation
cross sections required to explain recent data, through the ``Sommerfeld enhancement" that increases
the cross section at low velocities \footnote{Another possibility for sufficiently light bosons would be radiative capture \cite{Pospelov:2008jd}.}. A simple classical analogy can be used to illustrate the effect.
Consider a point particle impinging on a star of radius $R$.
Neglecting gravity, the cross section for the
particle to hit and be absorbed by the star is $\sigma_0 = \pi R^2$.
However, because of gravity, a point coming from a larger impact
parameter than $R$ will be sucked into the star. The cross section
is actually $\sigma = \pi b_{max}^2$, where $b_{max}$ is the largest
the impact parameter can be so that the distance of closest approach
of the orbit is $R$. If the velocity of the particle at
infinity is $v$, we can determine $b_{max}$ trivially using
conservation of energy and angular momentum, and we find that
\begin{equation}
\sigma = \sigma_0 \left(1 + \frac{v_{esc}^2}{v^2}\right)
\end{equation}
where $v^2_{esc} = 2 G_N M/R$ is the escape velocity from the surface
of the star.
For $v \ll v_{esc}$, there is a large enhancement of
the cross section due to gravity; even though the correction
vanishes as gravity shuts off ($G_N \to 0$), the expansion parameter
is $2 G_N M/(R v^2)$ which can become large at small velocity.

The Sommerfeld enhancement is a  quantum counterpart to this
classical phenomenon. It  arises whenever a particle has
an attractive force carrier with a Compton
wavelength longer than $(\alpha M_{DM})^{-1}$, i.e. dark matter bound states are
present in the spectrum of the theory. (We generically refer to $\alpha \sim {\rm coupling}^2/4 \pi$, assuming such couplings are comparable to those in the standard model, with $10^{-3} \lsim \alpha \lsim 10^{-1}$.)

Let us study this enhancement more quantitatively. We begin with the simplest case of interest, namely, a particle interacting via a Yukawa potential. We assume a dark matter particle $\chi$ coupling to a mediator $\phi$ with coupling strength $\lambda$.
For $s$-wave annihilation in the nonrelativistic limit, the reduced
two-body wavefunction obeys the radial Schr\"{o}dinger equation,
\begin{equation} \frac{1}{m_{\chi}} \psi''(r) - V(r) \psi(r) = -m_{\chi} v^2\psi(r) , \label{schrodinger} \end{equation}
where the $s$-wave wavefunction $\Psi(r)$ is related to $\psi(r)$ as $\Psi(r)=\psi(r)/r$, $v$ is the velocity of each particle in the center-of-mass frame (here we use units where $\hbar=c=1$), and for scalar $\phi$ the potential takes the usual Yukawa form,
\begin{equation} V(r) = -\frac{\lambda^2}{4 \pi r} e^{- m_{\phi} r}. \end{equation}
The interaction in the absence of the potential is pointlike. As
reviewed in the appendix, the Sommerfeld enhancement in the scattering
cross section due to the potential is given by
\begin{equation}
S = \left|\frac{\frac{d \psi_k}{dr}(0)}{k}\right|^2
\end{equation}
 where we solve the Schr\"{o}dinger equation with boundary conditions $\psi(0) = 0$, $\psi(r) \to \sin(k r + \delta)$ as $r \to \infty$. In the recent dark matter literature, a different but completely equivalent expression is used, with
 \begin{equation}
 S = |\psi(\infty)/\psi(0)|^2
 \end{equation}
 where we solve the Schr\"{o}dinger equation with the outgoing boundary condition $\psi'(\infty) = i m_{\chi} v \psi(\infty)$ \cite{2005PhRvD..71f3528H}.

Defining the dimensionless parameters
\begin{equation}
\alpha = \lambda^2 / 4 \pi, \,\, \epsilon_v= \frac{v}{\alpha}, \,\, \epsilon_\phi = \frac{m_{\phi}}{\alpha m_{\chi}},
\end{equation}
and rescaling the radial coordinate with $r' = \alpha m_{\chi} r$, we can rewrite Eq. \ref{schrodinger} as,
\begin{equation} \psi''(r') + \left(\epsilon_v^2 + \frac{1}{r'} e^{- \epsilon_\phi r'} \right)\psi(r') = 0. \label{dimensionless_SE} \end{equation}

In the limit where the $\phi$ mass goes to zero ($\epsilon_\phi \rightarrow 0$), the effective potential is just the Coulomb potential and Eq. \ref{dimensionless_SE} can be solved analytically, yielding an enhancement factor of,
\begin{equation} S \equiv |\psi(\infty)/\psi(0)|^2 = \frac{\pi/\epsilon_v}{1 - e^{-\pi/\epsilon_v}}. \label{massless_enhancement} \end{equation}
For nonzero $m_\phi$ and hence nonzero $\epsilon_\phi$,  there are
two important qualitative differences. The first is that the
Sommerfeld enhancement saturates at low velocity--the attractive
force has a finite range, and this limits how big the enhancement
can get. At low velocities, once the deBroglie wavelength of the
particle $(Mv)^{-1}$ gets larger than the range of the interaction
$m^{-1}_\phi$, or equivalently once $\epsilon_v$ drops beneath
$\epsilon_\phi$, the Sommerfeld enhancement saturates at $S \sim
\frac{1}{\epsilon_\phi}$. The second effect is that for specific
values of  $\epsilon_\phi$, the Yukawa potential develops threshold
bound states, and these give rise to resonant enhancements of the
Sommerfeld enhancement. We describe some of the parametrics for
these effects in the appendix, but for reliable numbers Eq.
\ref{dimensionless_SE} must be solved numerically, and plots for the
enhancement as a function of $\epsilon_\phi$ and $\epsilon_v$ are
given there. As we will see in the following, we will be interested
in a range of $m_\phi \sim 100$ MeV - GeV; with reasonable values of
$\alpha$, this corresponds to $\epsilon_\phi$ in the range $\sim
10^{-2} - 10^{-1}$, yielding Sommerfeld enhancements ranging up
to a factor $\sim 10^3 - 10^4$. At low velocities, the finite range of the Yukawa interaction causes the Sommerfeld enhancement to saturate, so the enhancement factor cannot greatly exceed this value even at arbitrarily low velocities. The nonzero mass of the $\phi$ thus prevents the catastrophic overproduction of gammas in
the early universe pointed out by \cite{Kamionkowski:2008gj}. 

\begin{figure}[ht]
\centering
\includegraphics[width=0.5\textwidth]{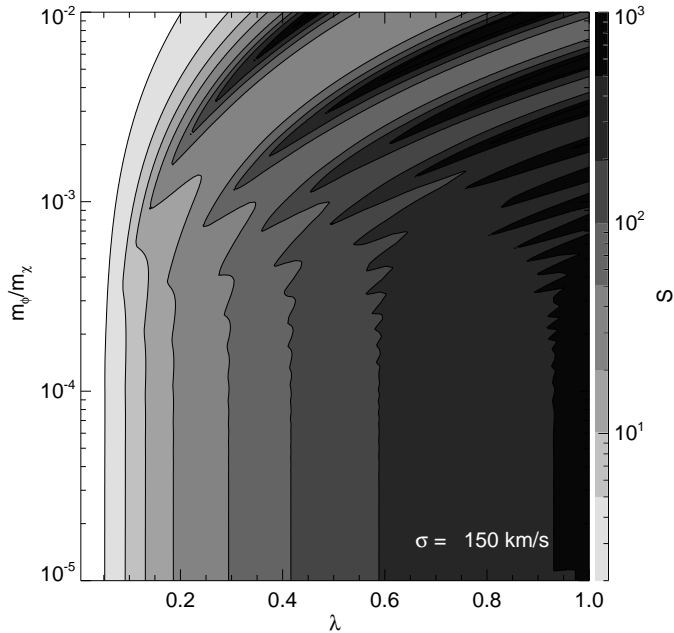}
\caption{Contours for the Sommerfeld enhancement factor $S$ as a function of the mass ratio $m_{\phi}/m_{\chi}$ and the coupling constant $\lambda$, $\sigma = 150$ km/s.}
\label{v=150}
\end{figure}

Having obtained the enhancement $S$ as a function of $\epsilon_v$
and $\epsilon_\phi$, we must integrate over the velocity
distribution of the dark matter in Earth's neighborhood, to obtain
the total enhancement to the annihilation cross section for a
particular choice of $\phi$ mass and coupling $\lambda$. We assume a
Maxwell-Boltzmann distribution for the one-particle velocity,
truncated at the escape velocity:
\begin{equation} f(v) = \left\{ \begin{array}{cc} N v^2e^{-v^2 / 2 \sigma^2} & v \leq v_{\text{esc}} \\ 0 & v > v_{\text{esc}}. \end{array} \right. \end{equation}
The truncation does not significantly affect the results, as the enhancement factor drops rapidly with increasing velocity. The one-particle rms velocity is taken to be 150 km/s in the baseline case, following simulations by Governato et al. \cite{Governato:2007}. Fig. \ref{v=150} shows the total enhancement as a function of $m_{\phi}/m_{\chi}$ and the coupling $\lambda$ for this case.

\begin{figure}[ht]
\centering
\includegraphics[width=0.5\textwidth]{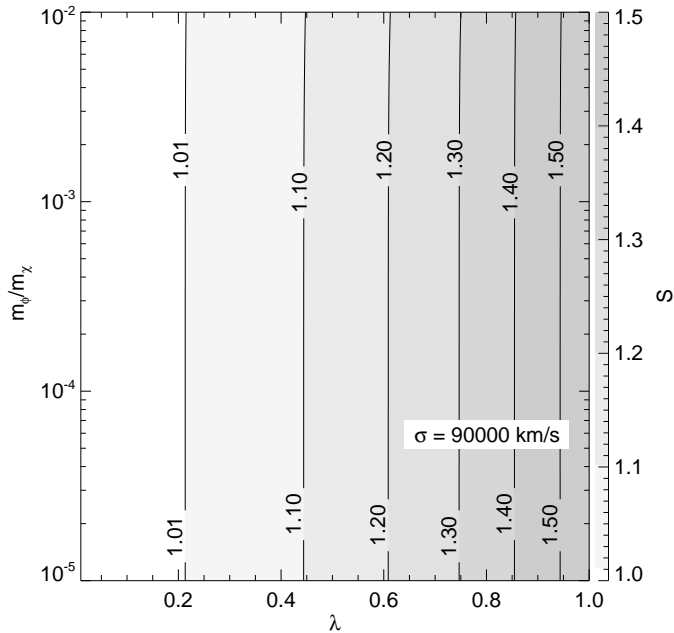}
\caption{Contours for the Sommerfeld enhancement factor $S$ as a function of the mass ratio $m_{\phi}/m_{\chi}$ and the coupling constant $\lambda$, at a temperature $T_{\rm{CMB}}=m_\chi/20$ (the enhancement is integrated over a Boltzmann distribution with $\sigma = 0.3 c$).}
\label{earlyuniverse}
\end{figure}

One can inquire as to whether dark matter annihilations in the early universe experience the same
Sommerfeld enhancement as dark matter annihilations in the galactic halo at the present time.  This is important, because we are relying on
this effect precisely to provide us with an annihilation cross section
in the present day much larger than that in the early universe. However, it turns out that particles leave thermal equilibrium long
before the Sommerfeld enhancement turns on. This is because the
Sommerfeld enhancement occurs when the expansion parameter $\alpha/v =
1 / \epsilon_v$ is large. In the early universe, the dark matter
typically decouples at $T_{\rm{CMB}} \sim m_\chi/20$, or $v \sim 0.3
c$. Since we are taking $\alpha \lsim 0.1$ generally, the Sommerfeld
enhancement has not turned on yet. More precisely, in the
early-universe regime $\epsilon_v \gg \epsilon_\phi$, so we can use
Eq. \ref{massless_enhancement} for the massless limit. Where
$\epsilon_v \gg 1$, Eq. \ref{massless_enhancement} yields $S \approx 1
+ \pi / 2 \epsilon_v$: thus the enhancement should be small and
independent of $m_\phi$. Fig. \ref{earlyuniverse} shows this
explicitly. We are left with the perturbative annihilation cross
section $\sigma \sim \alpha^2/m_\chi^2$ which gives us the usual
successful thermal relic abundance.

At some later time, as the dark matter velocities redshift to
lower values, the Sommerfeld enhancement turns on and the
annihilations begin to scale as $a^{-5/2}$ (before kinetic decoupling)
or $a^{-2}$ (after decoupling). From decoupling until matter-radiation
equality, where the
Hubble scale begins to evolve differently, or until the Sommerfeld
effect is saturated, dark matter annihilation will produce a uniform
amount of energy per comoving volume per Hubble time. This uniform
spread of energy injected could have potentially interesting signals
for observations of the early universe. An obvious example would be a
possible effect on the polarization of the CMB, as described in
\cite{Peebles:2000pn,Padmanabhan:2005es,Mapelli:2006ej}. Because at the time of matter-radiation
equality, the dark matter may have slowed to velocities of $v \sim
10^{-6} c$ or slower, the large cross sections could yield a
promising signal for upcoming CMB polarization observations, including
\emph{Planck}.  However, we emphasize that saturation of the cross
section at low $v$ avoids the runaway annihilations discussed by
\cite{Kamionkowski:2008gj}.

\section{Models of the Sommerfeld Force and New Annihilation Channels}
What sorts of forces could give rise to a large Sommerfeld enhancement of the dark matter annihilation? As we have already discussed, we must have a {\em light} force carrier. On the other hand, a massless particle is disfavored by the agreement between big bang nucleosynthesis calculations and primordial light element measurements \cite{Steigman:2007xt}, as well as constraints from WMAP on new relativistic degrees of freedom \cite{Hinshaw:2008kr}. Thus, we must have massive degrees of freedom, which are naturally light, while still coupling significantly to the dark matter. There are three basic candidates.

\begin{itemize}
\item The simplest possibility is coupling to a light scalar field, which
does give rise to an attractive interaction. However, given that we
need an $\mathcal{O}(1)$ coupling to the DM fields, this will typically make
it unnatural for the scalar to stay as light as is needed to
maximize the Sommerfeld enhancement, unless the dark matter sector
is very supersymmetric. This can be a challenge given that we are expecting the dark matter to have a mass $\mathcal{O}$(TeV). Consequently, the natural scale for a scalar which couples to it would also be $\mathcal{O}$(TeV), although this conclusion can be evaded with some simple model-building.

\item  The scalar could be naturally light if it is a pseudoscalar
$\pi$ with a goldstone-like derivative coupling to matter $1/F J_\mu
\partial^\mu \pi$. This does lead to a long-range spin-dependent potential of the
form $V(\vec{r}) = \frac{1}{r^3}(\vec{S}_1 \cdot \vec{S}_2 - 3
\vec{S}_1 \cdot \hat{r} \vec{S}_2 \cdot \hat{r})$, but the numerator
vanishes when averaged over angles, so there is no long-range
interaction in the $s$-wave and hence no Sommerfeld enhancement.

\item Finally, we can have a coupling to spin-1 gauge fields arising from some dark gauge symmetry $G_{\rm dark}$. Since the gauge fields must have a mass $\mathcal{O}$(GeV) or less, one might worry that this simply begs the question, as the usual explanation of such a light gauge boson requires the existence of a scalar with a mass of $\mathcal{O}$(GeV) or less. However, because that scalar need not couple directly to the dark matter, it is sufficiently sequestered that its small mass is technically natural. Indeed, the most straightforward embedding of this scenario within SUSY \cite{nimaneal} naturally predicts the breaking scale for $G_{\rm dark}$ near $\sim$ GeV. Alternatively, no fundamental scalar is needed, with the vector boson possibly generated by the condensate of a strongly coupled theory.

\end{itemize}

At this juncture it is worth discussing an important point. As we
have emphasized, to produce a Sommerfeld {\it enhancement}, we need
an attractive interaction. Scalars (like gravitons) universally
mediate attractive forces, but gauge fields can give attraction or
repulsion, so do we generically get a Sommerfeld enhancement? For
the case of the Majorana fermion (or real scalar) in particular, the
dark matter does not carry any charge, and there does not appear to
be any long-range force to speak of, so the question of the
enhancement is more interesting. As we discuss in a little more
detail in the appendix, the point is that the gauge symmetry is
broken. The breaking dominates the properties of the asymptotic
states. For instance, the dark matter must be part of a multiplet
with at least two states, since a  spin-1 particle cannot have a
coupling to a single neutral state. The gauge symmetry breaking
leads to a mass splitting between the states, which dominates the
long-distance behavior of the theory, determining which state is the
lightest, and therefore able to survive to the present time and
serve as initial states for collisions leading to annihilation.
However, if the mass splitting between the states is small enough
compared to the kinetic energy of the collision, the gauge-partner
DM states will necessarily be active in the collision, and
eventually at distances smaller than the gauge boson masses, the
gauge breaking is a negligible effect. Since the asymptotic states
are in general roughly equal linear combinations of positive and
negative charge gauge eigenstates, the remnant of asymptotic
gauge breaking at short distances is that the incoming scattering
states are linear combinations of gauge eigenstates, so the two-body
wavefunction will be a linear combination of attractive and
repulsive channels. While two-body wavefunction in the repulsive
channel is indeed suppressed at the origin, the attractive part is
enhanced. Therefore there is  still a Sommerfeld enhancement,
suppressed at most by an $\mathcal{O}(1)$ factor reflecting the
attractive component of the two-body state. (This is of course
completely consistent with the Sommerfeld enhancements seen for
ordinary WIMP annihilations, mediated by W/Z/$\gamma$ exchange).

\begin{figure}
\includegraphics[width=0.8\textwidth]{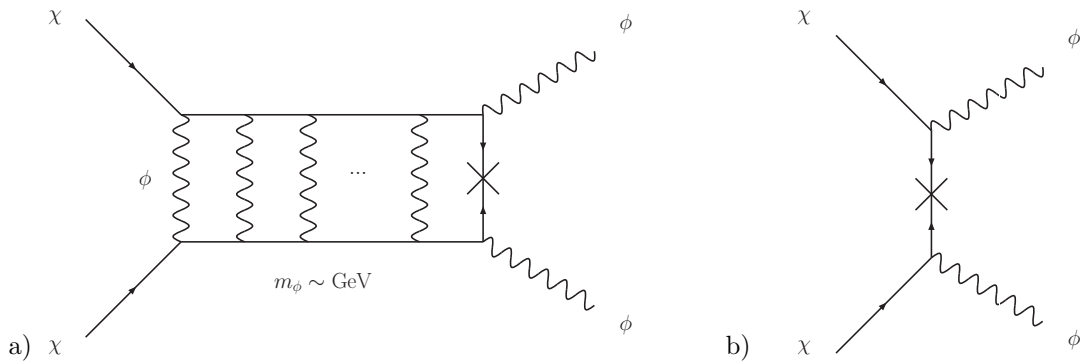}
\caption{The annihilation diagrams $\chi \chi \rightarrow \phi \phi$ both with (a) and without (b) the Sommerfeld enhancements.}
\label{fig:annihilation}
\end{figure}

Because of the presence of a new light state, the annihilation $\chi
\chi \rightarrow \phi \phi$ can, and naturally will, be significant.
In order not to spoil the success of nucleosynthesis, we cannot have
very light new states in this sector, with a mass $\lsim 10$ MeV, in
thermal equilibrium with the standard model; the simplest picture is
therefore that all the light states in the dark sector have a mass
$\sim$ GeV. Without any special symmetries, there is no reason for
any of these particles to be exactly stable, and the lightest ones
can therefore only decay back to standard model states, indeed many
SM states are also likely kinematically inaccessible, thus favoring
ones that produce high energy positrons and electrons. This
mechanism was first utilized in \cite{Cholis:2008vb} to generate a
large positron signal with smaller $\pi^0$ and $\bar p$ signals.
Consequently, an important question is the tendency of $\phi$ to
decay to leptons. This is a simple matter of how $\phi$ couples to
the standard model. (A more detailed discussion of this can be found
in \cite{nimaneal}.)

A scalar $\phi$ can couple with a dilaton-like coupling $\phi F^{\mu \nu} F_{\mu \nu}$, which will produce photons and hadrons (via gluons). Such a possibility will generally fail to produce a hard $e^+e^-$ spectrum. A more promising approach would be to mix $\phi$ with the standard model Higgs with a term $\kappa \phi^2 h^\dagger h$. Should $\phi$ acquire a vev $\vev{\phi} \sim m_\phi$, then we yield a small mixing with the standard model Higgs, and the $\phi$ will decay into the heaviest fermion pair available. For $m_\phi \lsim 200 \mev$ it will decay directly to $e^+e^-$, while for $200 \mev \lsim m_\phi \lsim 250 \mev$, $\phi$ will decay dominantly to muons. Above that hadronic states appear, and pion modes will dominate. Both $e^+e^-$ and $\mu^+\mu^-$ give good fits to the PAMELA data, while $e^+e^-$ gives a better fit to PAMELA+ATIC.

A pseudoscalar, while not yielding a Sommerfeld enhancement, could naturally be present in this new sector. Such a particle would typically couple to the heaviest particle available, or through the axion analog of the dilaton coupling above. Consequently, the decays of a pseudoscalar would be similar to those of the scalar.

A vector boson will naturally mix with electromagnetism via the operator $F_{\mu \nu}' F^{\mu \nu}$. This possibility was considered some time ago in \cite{Holdom:1985ag}. Such an operator will cause a vector $\phi_\mu$ to couple directly to charge. Thus, for $m_\phi \lsim 2 m_\mu$ it will decay to $e^+e^-$, while for $2 m_\mu \lsim m_\phi \lsim 2 m_\pi$ it will decay equally to $e^+e^-$ and $\mu^+ \mu^-$. Above $2 m_\pi$, it will decay $40\%\, e^+e^-$, $40\%\, \mu^+ \mu^-$ and $20 \%\, \pi^+ \pi^-$. At these masses, no direct decays into $\pi^0$'s will occur because they are neutral and the hadrons are the appropriate degrees of freedom. At higher masses, where quarks and QCD are the appropriate degrees of freedom, the $\phi$ will decay to quarks, producing a wider range of hadronic states, including $\pi^0$'s, and, at suitably high masses $m_\phi \gsim 2 \gev$, antiprotons as well \footnote{Recently, \cite{Barger:2008su} explored the possibility that PAMELA could distinguish the spin of the dark matter, which could be relevant here in annihilations to vector bosons.}. In addition to XDM \cite{Finkbeiner:2007kk}, some other important examples of theories under which dark matter interacts with new forces include WIMPless models \cite{Feng:2008ya}, mirror dark matter \cite{Foot:1995pa} and secluded dark matter \cite{Pospelov:2007mp}.

Note that, while these interactions between the sectors can be small, they are all large enough to keep the dark and standard model sectors in thermal equilibrium down to temperatures far beneath the dark matter mass, and (as mentioned in the previous section), we can naturally get the correct thermal relic abundance with a weak-scale dark matter mass and perturbative annihilation cross sections. Kinetic equilibrium in these models is naturally maintained down to the temperature $T_{\rm{CMB}} \sim m_\phi$ \cite{2008arXiv0805.3531F}.

\section{A Non-Abelian $G_{\rm dark}$: INTEGRAL, Direct Detection, and DAMA}
Up to this point we have focused on a situation where there is a
single force-carrying boson $\phi$, whether vector or scalar. Already,
this can have significant phenomenological consequences. In mixing
with the standard model Higgs boson, there is a nuclear recoil cross
section mediated by $\phi$. With technically natural parameters as
described in \cite{Finkbeiner:2007kk}, the rate is unobservable,
although in a two-Higgs doublet model the cross section is within
reach of future experiments \cite{tracy}.

In contrast, an $800\gev$ WIMP which interacts via a particle that couples to charge is strongly constrained. Because the $\phi$ boson is light and couples to the electromagnetic vector current, there are strong limits. The cross section per nucleon for such a particle is \cite{Pospelov:2007mp}
\be
\sigma_0 &=& \frac{16 \pi  Z^2 \alpha_{\rm{SM}} \alpha_{\rm{Dark}} \epsilon^2 \mu_{\rm{ne}}^2}{A^2 m_\phi^4} \\ \nonumber &=& \left(\frac{Z}{32}\right)^2\left(\frac{73}{A}\right)^2 \left(\frac{\epsilon}{10^{-3}} \right)^2 \left( \frac{\alpha_{\rm{Dark}}}{137^{-1}}\right) \left(\frac{\mu_{\rm{ne}}}{938\mev}\right)\left(\frac{1\, \gev}{m_\phi}\right)^4 \times 1.8 \times 10^{-37} \cm^2,
\ee
where $\alpha_{\rm{Dark}}$ is the coupling of the $\phi$ to the dark matter, $\epsilon$ describes the kinetic mixing, $\mu_{\rm{ne}}$ is the reduced mass of the DM-nucleon system and $\alpha_{\rm{SM}}$ is the standard model electromagnetic coupling constant. With the parameters above, such a scattering cross section is excluded by the present CDMS \cite{Ahmed:2008eu} and XENON \cite{Angle:2007uj} bounds by 6 orders of magnitude. However, this limit can be evaded by splitting the two Majorana components of the Dirac fermion \cite{Smith:2001hy} or by splitting the scalar and pseudoscalar components of a complex scalar \cite{Han:1997wn,Hall:1997ah}. Since the vector coupling is off-diagonal between these states, the nuclear recoil can only occur if there is sufficient kinetic energy to do so. If the splitting $\delta > v^2 \mu/2$ (where $\mu$ is the reduced mass of the WIMP-nucleus system) no scattering will occur.
Such a splitting can easily arise for a $U(1)$ symmetry by a $U(1)$ breaking operator such as
$
\frac{1}{M}\psi^c \psi h^* h^*
$
which generates a small Majorana mass and splits the two components (see \cite{Tucker-Smith:2004jv} for a discussion).

Remarkably, for $\delta \sim 100 \kev$ one can reconcile the DAMA
annual modulation signature with the null results of other experiments
\cite{Smith:2001hy, Tucker-Smith:2004jv, Chang:2008gd}, in the
``inelastic dark matter'' scenario. We find that the ingredients
for such a scenario occur here quite naturally.
However, the splitting here must be $\mathcal{O}(100\kev)$ and the origin of this scale is unknown, a point we shall address shortly.

\vskip 0.2in
{\noindent \bf Exciting Dark Matter from a Non-Abelian Symmetry}

\vskip 0.1in

One of the strongest motivations for a $\sim \gev$ mass $\phi$ particle prior to the present ATIC and PAMELA
data was in the context of eXciting dark matter \cite{Finkbeiner:2007kk}.
In this scenario, dark matter excitations could occur in the center of the galaxy via
inelastic scattering $\chi \chi \rightarrow \chi \chi^*$.
If $\delta = m_{\chi^*} - m_\chi \gsim  2 m_e$, the decay $\chi^* \rightarrow \chi e^+ e^-$
can generate the excess of $511 \, \kev$ x-rays seen from the galactic center by the
INTEGRAL \cite{Weidenspointner:2006nu,Weidenspointner:2007rs} satellite.
However, a large (nearly geometric) cross section is needed to produce the large numbers of positrons
 observed in the galactic center, necessitating a boson with mass of the order of the momentum transfer, i.e.,
 $m_\phi \lsim M_\chi v \sim \gev$, precisely the same scale as we require for the Sommerfeld enhancement.
 But where does the scale $\delta \sim \mev$ come from? Remarkably, it arises radiatively at precisely the appropriate scale\footnote{Note that in the early universe, all these states will be populated at freeze-out, and will participate in maintaining thermal equilibrium. However, unlike coannihilation in the MSSM, which typically occurs between a Bino and a slepton, there is not expected to be a large effect, on the connection between the present annihilation rate and that at freeze-out. This is because the self-annihilation and coannihilation cross sections are similar or even equal here, while in the MSSM the coannihilation cross section is much larger.}.

We need the dark matter to have an excited state, and we will
further assume the dark matter transforms under a non-Abelian gauge
symmetry. Although an excited state can be present with simply a
$U(1)$, this only mediates the process $\chi \chi \rightarrow \chi^*
\chi^*$. If this requires energy greater than $4 m_e$ it is very
hard to generate enough positrons to explain the INTEGRAL signal. If
we assume the dark matter is a Majorana fermion, then it must
transform as a real representation of the gauge symmetry. For a
non-Abelian symmetry, the smallest such representation will be
three-dimensional [such as a triplet of SU(2)]. This will allow a
scattering $\chi_1 \chi_1 \rightarrow \chi_2 \chi_3$. If $m_3$ is
split from  $m_2 \sim m_1$ by an amount $\delta \sim \mev$, we have
arrived at the setup for the XDM explanation of the INTEGRAL signal.

Because the gauge symmetry is Higgsed, we should expect a splitting
between different states in the dark matter multiplet. This could
arise already at tree-level, if the dark matter has direct couplings
to the Higgs fields breaking the gauge symmetry; these could naturally be
as large as the dark gauge breaking scale $\sim$ GeV itself, which
would be highly undesirable, since we need these splittings to be
not much larger than the DM kinetic energies in order to get a
Sommerfeld enhancement to begin with. However, such direct couplings
to the Higgs could be absent or very small (indeed most of the
Yukawa couplings in the standard model are very small). We will
assume that such a direct coupling is absent or negligible. The
gauge breaking in the gauge boson masses then leads, at one loop, to
splittings between different dark matter states, analogous to the
familiar splitting between charged and neutral components of a
Higgsino or Dirac neutrino. These splittings arise from infrared
effects, and so are completely calculable, with sizes generically
$\mathcal{O}(\alpha m_Z)$ in the standard model or
$\mathcal{O}(\alpha m_\phi)\sim \mev$ in the case at hand. Thus we
find that the MeV splittings needed for XDM arise {\em
automatically} once the mass scale of the $\phi$ has been set to
$\mathcal{O}(GeV)$.

\begin{figure*}
\begin{center}
\includegraphics[width=0.5\textwidth]{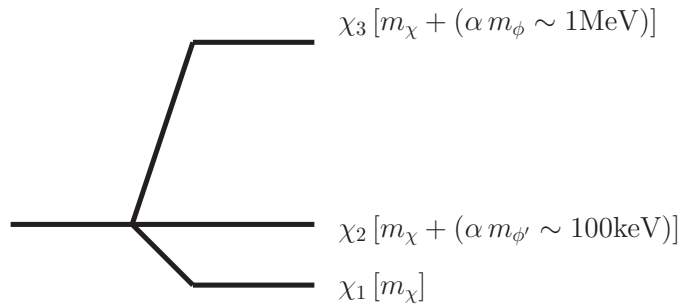}
\end{center}
\caption{Spectrum of exciting dark matter.}
\label{fig:spectrum}
\end{figure*}

We would like $\chi_2$ and $\chi_1$ to stay similar in mass, which can occur if the breaking pattern approximately preserves a custodial symmetry. However, if they are too degenerate, we are forced to take $\epsilon < 10^{-5}$ in order to escape direct-detection constraints. On the other hand, we do not want {\em too} large of a splitting between $\chi_2$ and $\chi_1$, as this would suppress the rate of positron production for INTEGRAL. Thus, we are compelled to consider $\delta_{21} \sim 100-200 \kev$, which puts us precisely in the range relevant for the inelastic dark matter explanation of DAMA. (See Fig. \ref{fig:spectrum}.)

All of these issues require detailed model-building, which we defer to future work; however, existence proofs are easy to construct. The Lagrangian is of the form
\begin{equation}
{\cal L} = {\cal L}_{\text{SM}} + {\cal L}_{\text{Dark}} + {\cal L}_{\text{mix}}.
\end{equation}
As a familiar example, imagine that $G_{\text{Dark}} = SU(2) \times U(1)$, with gauge bosons $w_{\mu I}$ and $b_\mu$ which we collectively refer to as $a_{\mu i}$, and
the dark matter multiplet $\chi$ transforming as a triplet under $SU(2)$ and neutral under the $U(1)$; it could be either a scalar or fermion.  We also assume that some set of Higgses completely break the symmetry.
Working in unitary gauge, the tree-level dark sector Lagrangian is
\begin{equation}
{\cal L}_{\text{Dark}} = {\cal L}_{\text{Gauge Kin.}} +
\frac{1}{2} m^2_{ij} a^\mu_i a_{\mu j} + \cdots
\end{equation}
where the $m^2_{ij}$ makes all the dark spin-1 fields massive, and $\cdots$ refers to other fields such as the physical Higgses that could be present.
At one loop, this broken gauge symmetry will induce splittings between the 3 real DM states all of $\mathcal{O}(\alpha_{Dark} m_{Dark})$ as just discussed above.
The leading interaction between the two sectors is via kinetic mixing between the new $U(1)$ and the photon (which is inherited from such a mixing with hypercharge):
\begin{equation}
{\cal L}_{\text{mix}} = \frac{1}{2} \epsilon  b_{\mu \nu} F^{\mu \nu}
\end{equation}
We put an $\epsilon$ in front of this coupling because it is natural
for this coupling to be small; it can be induced at one loop by
integrating out some heavy states (of any mass between the GeV and
Planck scales) charged under both this new $U(1)$ and hypercharge.
This can easily make $\epsilon \sim 10^{-4} - 10^{-3}$. Even without
a $U(1)$, a similar mixing could be achieved with an ``S-parameter''
type operator Tr[$(\Phi/M)^p G_{\mu \nu}] F_{\mu \nu}$, where $\Phi$
is a dark Higgs field with quantum numbers such that $(\Phi/M)^p$
transforms as an adjoint under $G_{\text{Dark}}$; it is reasonable to
imagine that the scale $M$ suppressing this operator is near the
weak scale.

Going back to PAMELA/ATIC, the non-Abelian self-couplings of the
vector bosons can have an interesting effect on the annihilation
process. For large enough $\alpha_{\text{Dark}}$, the gauge bosons radiate
other soft and collinear gauge bosons leading to a ``shower"; this
can happen when $\alpha_{\rm{Dark}} \log^2(M_\chi/m_\phi) \gsim 1$. While for
quite perturbative values of $\alpha_{\text{Dark}}$ this is not an
important effect, it could be interesting for larger values, and
would lead to a greater multiplicity of softer $e^+ e^-$ pairs in
the final state.

Before closing this section, it is important to point out that it is
not merely a numerical accident that the excited states relevant for
INTEGRAL and DAMA can actually be excited in the DM-DM collisions
and DM collisions with direct-detection nuclei, but is rather a
parametric consequence of maximizing the Sommerfeld enhancement for
the annihilation cross section needed to explain ATIC/PAMELA. As we
already emphasized in our discussion of Sommerfeld enhancement with
vector states, vector boson couplings necessarily connect different
dark matter mass eigenstates, and therefore there is no enhancement
for the annihilation cross section needed to explain ATIC/PAMELA if
the mass splittings are much larger than the kinetic energy
available in the collision. But this parametrically implies that
dark matter collisions should also have the kinetic energy needed to
create the excited states, as necessary for the INTEGRAL signal. It
is also interesting to note that the condition needed for the large
geometric capture cross section, $m_\phi \lsim M v$, also tells us
that the Sommerfeld enhancement is as large as can be at these
velocities, and has not yet been saturated by the finite range of
the force carrier. Furthermore, since the mass of the heavy nuclei
in direct detection experiments is comparable to the dark matter
masses, the WIMP-nucleus kinetic energy is also naturally comparable
to the excited state splittings. In this sense, even absent the
direct experimental hints, signals like those of INTEGRAL and
inelastic scattering for direct detection of dark matter are
parametric predictions of our picture.

\section{Substructure and the Sommerfeld Enhancement}
As we have seen, the Sommerfeld enhancement leads to a cross section that scales at low energies as $\sigma v \sim 1/v$. This results in a relatively higher contribution to the dark matter annihilation from low velocity particles. While the largest part of our halo is composed of dark matter particles with an approximately thermal distribution, there are subhalos with comparable or higher densities. Because these structures generally have lower velocity dispersions than the approximately thermal bulk of the halo, the Sommerfeld-enhanced cross sections can make these components especially important.

\begin{figure*}
\centering
\includegraphics[width=0.5\textwidth]{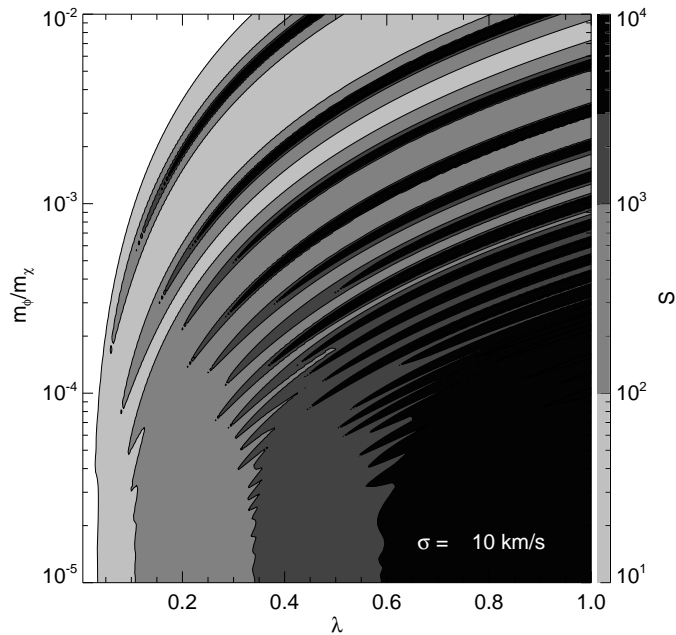}
\caption{Contours for the Sommerfeld enhancement factor $S$ as a function of the mass ratio $m_{\phi}/m_{\chi}$ and the coupling constant $\lambda$, $\sigma = 10$ km/s.}
\label{v=10}
\end{figure*}

Subhalos of the Milky Way halo are of particular interest, and N-body
simulations predict that many should be present. There is still some debate as to what effect
substructures can have on indirect detection prospects. However,
some of these subhalos will have velocity dispersions of order
$10~\kms$ \cite{Strigari:2006rd}, and a simple examination of Fig. \ref{v=10} shows that dramatic increases of up to two orders of magnitude in the Sommerfeld enhancement can occur for these lower velocities!

Although in our model there are no direct $\pi^0$ gammas, there are still significant hopes for detection of dwarf galaxies through dark matter annihilation. If the DM is also charged under $SU(2)\times U(1)$ there may be a subdominant component of annihilations into $W^+W^-$ for instance, much larger than the $s$-wave limited thermal cross section, which could yield significant photon signals from the hadronic shower\footnote{Although larger annihilation rates into $W^+W^-$ or other standard model states are possible, the cross section must be lower than the limits of \cite{Cirelli:2008pk,Donato:2008jk} arising from antiprotons. Moreover, strong limits from final state radiation \cite{Beacom:2004pe} can also arise, although these depend more sensitively on the nature of the halo profile.}
. The copious high energy positrons and electrons, produced even more abundantly than expected for a nonthermal WIMP generating the ATIC or PAMELA signal, can produce inverse Compton scattering signals off of the CMB. Even a loop suppressed annihilation into $\gamma \gamma$ may be detectable with this enhancement.

This presumes that the local annihilation does not have significant enhancement from a low-velocity component, however. If the local annihilation rate is also enhanced by substructure, then our expectations of the enhancement for dwarf galaxies would not be as large. As a concrete example, we can consider the possibility of a ``dark disk.'' Recently, it has been argued that the old stars in the thick Milky Way disk should have an associated dark matter component, with dynamics which mirror those stars, and a density comparable to the local density \cite{Read:2008fh}. Because of the low velocity dispersion (of order 10 \kms), our estimates of what is a reasonable Sommerfeld boost may be off by an order of magnitude. 

As a consequence of this and other substructure, the Sommerfeld boost of Fig. \ref{v=150} should be taken as a lower bound, with contributions from substructure likely increasing the cross section significantly further. We will not attempt to quantify these effects here beyond noting the possible increases of an order of magnitude or more suggested by Fig. \ref{v=10}. However, understanding them may be essential for understanding the size and spectrum if this enhancement of the cross section is responsible for the signals we observe.

\section{Outlook: Implications for PAMELA, Planck, Fermi/GLAST, and the LHC}
If the excesses in positrons and electrons seen by PAMELA and ATIC are
arising from dark matter, there are important implications for a wide
variety of experiments. It appears at this point that a simple
modification of a standard candidate such as a neutralino in the minimal supersymmetric standard model (MSSM) is
insufficient. The need for dominantly leptonic annihilation modes with
large cross sections significantly changes our intuition for what we
might see, where, and at what level. There are a few clear
consequences looking forward.

\begin{itemize}
\item{The positron fraction seen by PAMELA should continue to rise up to the highest energies available to them ($\sim 270\, \gev$), and the electron + positron signal should deviate from a simple power law, as seen by ATIC. }
\item{If the PAMELA/ATIC signal came from a local source, we would not expect additional anomalous electronic activity elsewhere in the galaxy. Dark matter, on the other hand, should produce a significant signal in the center of the galaxy as well, yielding significant signals in the microwave range through synchrotron radiation and in gamma rays through inverse-Compton scattering. The former may already have been seen at WMAP \cite{Finkbeiner:2003im,Hooper:2007kb} and the latter at EGRET \cite{strong:2005,thompson:2005}. Additional data from Planck and Fermi/GLAST will make these signals robust \cite{WIP}.}
\item{We have argued that the most natural way to generate such a large signal is through the presence of a new, light state, which decays dominantly to leptons. It is likely these states could naturally be produced at the LHC in some cascade, leading to highly boosted pairs of leptons as a generic signature of this scenario \cite{nimaneal}.}
\item{Although the cross section is not Sommerfeld enhanced during freeze-out, it can keep pace with the expansion rate over large periods of the cosmic history, between kinetic decoupling and matter-radiation equality. This can have significant implications for a variety of early-universe phenomena as well as the cosmic gamma-ray background \cite{WIP}.}
\item{The Sommerfeld enhancement is increasingly important at low velocity. Because substructure in the halo typically has velocity dispersions an order of magnitude smaller than the bulk of the halo, annihilations can be an order of magnitude higher, or more. With the already high local cross sections, this makes the prospects for detecting substructure even higher. With the mass range in question, continuum photons would be possibly visible at GLAST, while monochromatic photons [which could be generated in some models \cite{nimaneal} would be accessible to air Cerenkov telescopes, such as HESS (see \cite{Mack:2008wu} for a discussion, and \cite{2008arXiv0809.3894H} for HESS limits on DM annihilation in the Canis Major overdensity)].}

\end{itemize}

We have argued that dark matter physics is far richer than
usually thought, involving a multiplet of states and a new sector of
dark forces.  We have been led to propose this picture not by a
flight of fancy but rather directly from experimental data. Even so,
one can justifiably ask whether such extravagances are warranted.
After all, experimental anomalies come and go, and it is entirely
possible that the suite of hints that motivate our proposal are
incorrect, or that they have more conventional explanations.
However, we are very encouraged by the fact that the theory we have
presented fits into a very reasonable picture of particle physics,
is supported by overlapping pieces of experimental evidence, and
that features of the theory motivated by one set of experimental
anomalies automatically provide the ingredients to explain the
others. Our focus in this paper has been on outlining this unified
picture for dark matter; new experimental results coming soon should
be able to tell us whether these ideas are even qualitatively on the
right track. Needless to say it will then be important to find a
specific and simple version of this theory, with a small number of
parameters, to more quantitatively confront future data.

\vskip .1in

\noindent  {\bf Acknowledgements}

We would like to thank Ilias Cholis, Lisa Goodenough, Peter Graham,
Juan Maldacena, Patrick Meade, Michele Papucci, Nati Seiberg,
Leonardo Senatore, David Shih, Tomer Volansky, and Yosi Gelfand for
many stimulating discussions. We thank Matias Zaldarriaga for an
enjoyable discussion about classical and quantum Sommerfeld effects,
and David Shih for pointing out an error in our discussion of the
parametrics of the Sommerfeld effect for the Yukawa potential in v.1
of this paper, though our result is unchanged. We also thank Matt
Strassler, Patrick Meade, and Tomer Volansky for a discussion of
showering in annihilations to the non-Abelian $G_{Dark}$ gauge
bosons. The work of N.A.-H. is supported by the DOE under Grant
No. DE-FG02-91ER40654. NW is supported by NSF CAREER Grant No. PHY-0449818
and DOE OJI Grant No. DE-FG02-06ER41417.

\appendix

\section{A Quick Review of  Sommerfeld Enhancement}
The Sommerfeld enhancement is an elementary effect in nonrelativistic quantum mechanics; in this appendix we will review it in a simple way and discuss some of the parametrics for how the enhancement works for various kinds of interactions.

Consider a nonrelativistic particle moving
around some origin. There is an interaction Hamiltonian $H_{ann} =
U_{ann} \delta^3(\vec{r})$ localized to the origin, which e.g.
annihilates our particle or converts to another state in some way.
Imagine that the particle is moving in the $z$ direction so that its
wavefunction is
\begin{equation}
\psi_k^{(0)}(\vec{x}) = e^{i k z}
\end{equation}
then the rate for this process is proportional to
$|\psi^{(0)}(0)|^2$. But now suppose we also have a central
potential $V(r)$ attracting or repelling the particle to the origin.
We could of course treat $V$ perturbatively, but again at small
velocities the potential may not be a small perturbation and can
significantly distort the wave-function, which can be determined by
solving the Schr\"{o}dinger equation
\begin{equation}
-\frac{1}{2 M} \nabla^2 \psi_k + V(r) \psi_k = \frac{k^2}{2M} \psi_k
\end{equation}
with the boundary condition enforcing that the perturbation can only
produce outgoing spherical waves as $r \to \infty$:
\begin{equation}
\label{asymp} \psi \to e^{i k z} + f(\theta) \frac{e^{i k r}}{r} \,
{\rm as}  \, r \to \infty
\end{equation}
Now, since the annihilation is taking place locally near $r = 0$,
the only effect of the perturbation $V$ is to change the value of
the modulus of the wave-function at the origin relative to its
unperturbed value. Then, we can write
\begin{equation}
\sigma = \sigma_0 S_k
\end{equation}
where the Sommerfeld enhancement factor $S$ is simply
\begin{equation}
S_k = \frac{|\psi_k(0)|^2}{|\psi^{(0)}_k(0)|^2} = |\psi_k(0)|^2
\end{equation}
where we are using the normalization of the wavefunction $\psi_k$ as
given by the asymptotic form of Eq. \ref{asymp}.

 Finding a solution of the Schr\"{o}dinger equation with these asymptotics is a completely elementary and standard part of
scattering theory in nonrelativistic QM, which we quickly review
for the sake of completeness. Any solution of the Schr\"{o}dinger
equation with rotational invariance around the $z$ axis can be
expanded as
\begin{equation}
\psi_k = \sum_l A_l P_l(\cos \theta) R_{kl}(r)
\end{equation}
where $R_{kl}(r)$ are the continuum radial functions associated with
angular momentum $l$ satisfying
\begin{equation}
-\frac{1}{2M} \frac{1}{r^2} \frac{d}{dr} \left(r^2 \frac{d}{dr}
R_{kl}\right) + \left( \frac{l(l+1)}{r^2} + V(r) \right) R_{kl} = \frac{k^2}{2M} R_{kl}
\end{equation}
The $R_{kl}(r)$ are real, and at infinity look like a spherical
plane wave which we can choose to normalize as
\begin{equation}
R_{kl}(r) \to \frac{1}{r} \sin(kr - \frac{1}{2} l \pi + \delta_l(r))
\end{equation}
where $\delta_l(r) \ll kr$ as $ r \to \infty$.  The phase shift
$\delta_l(r)$ is determined by the requirement that $R_{kl}(r)$ is regular as $r \to 0$. Indeed, if the potential $V(r)$ does not blow up faster than $1/r$
near $r \to 0$, then we can ignore it relative to the kinetic terms,
and we have that $R_{kl}(r) \sim r^l$ as $r \to 0$; all but the
$l=0$ terms vanish at the origin. We now have to choose the coefficients
$A_l$ in order to ensure that the asymptotics of Eq. \ref{asymp}
are satisfied. Using the asymptotic expansion of $e^{i k z}$
\begin{equation}
e^{i k z} \to \frac{1}{2 i k r} \sum_l (2 l +1) P_l(\cos \theta)
\left[e^{i k r} - (-1)^l e^{- i k r} \right]
\end{equation}
determines the expansion to be
\begin{equation}
\psi_k = \frac{1}{k} \sum_l i^l (2 l +1) e^{i \delta_l} P_l(\cos
\theta) R_{kl}(r)
\end{equation}
It is now very simple to determine $\psi_k(0)$, since as we just
commented, $R_{k l}(r = 0)$ vanishes for every term other than
$l=0$. Thus, we have
\begin{equation}
S_k = \left|\frac{R_{k,l=0}(0)}{k}\right|^2
\end{equation}
We can furthermore make the standard substitution $R_{k,l=0} =
\chi_k/r$, then the Schr\"{o}dinger equation for $\chi$ turns into a
one-dimensional problem
\begin{equation}
\label{1dSch}
 -\frac{1}{2M} \frac{d^2}{dr^2} \chi_k + V(r) \chi_k =
\frac{k^2}{2M} \chi_k
\end{equation}
which we normalize at infinity with the condition
\begin{equation}
\label{1dasymp1} \chi_k(r) \to  \sin(kr + \delta)
\end{equation}
and since $R_{k,l=0}$ goes to a constant as $r \to 0$, we have to
have that $\chi \to 0$ as $r \to 0$, or
\begin{equation}
\label{1dasymp2} \chi_k(r) \to r \frac{d \chi_k}{dr}(0) \, \, {\rm
as \,} \, r \to 0
\end{equation}
Effectively, we can imagine launching $\chi$ from $\chi = 0$ at $r =
0$ with different velocities $\frac{d \chi_k}{dr}(0)$, and these
will evolve to some waveform as $r \to \infty$, but the correct
$\chi^\prime(0)$ is determined by the requirement that the waveform
at infinity have unit amplitude.

Summarizing, then, the Sommerfeld enhancement is
\begin{equation}
S_k = \left|\frac{\frac{d \chi_k}{dr}(0)}{k}\right|^2
\label{newenhance} \end{equation}
where $\chi_k$ satisfies the 1D Schr\"{o}dinger equation
\ref{1dSch} with boundary conditions
Eqs.\ref{1dasymp1}, \ref{1dasymp2}.  As a sanity check, let us see
how this works with vanishing potential. The solution that vanishes
as $r \to 0$ is $\chi_k(r) = A \sin(kr)$, and matching the
asymptotics forces $A = 1$. Then $\chi^\prime_k(0) = k$ and $S_k =
1$.

In the recent literature on the
subject,  a different expression for the Sommerfeld enhancement is used, arising from the
use of the optical theorem. We are instructed to solve the same 1D
Schr\"{o}dinger equation (Eq. \ref{1dSch}), this time with no special
boundary conditions at $\chi = 0$, but with boundary conditions so
that $\chi \propto e^{+i k r}$ as $r \to \infty$. Then, the
Sommerfeld enhancement is said to be
\begin{equation}
S_k = \frac{|\chi_k(\infty)|^2}{|\chi_k(0)|^2}
\label{oldenhance} \end{equation}
It is very easy to show that that these two
forms for $S_k$ agree exactly.
To see this, let us begin by denoting $\chi_1(r)$ to be the
solution to the Schr\"{o}dinger equation (Eq. \ref{1dSch}) with the boundary
condition $\chi_1(r) \rightarrow \sin(k r + \delta)$ as $r
\rightarrow \infty$ (Eq. \ref{1dasymp1}). As shown above, $\chi_1(0) = 0$. Let
$\chi_2(r)$ be the linearly independent solution with the boundary
condition $\chi_2(r) \rightarrow \cos(k r + \delta)$ as $r
\rightarrow \infty$, and define $A \equiv \chi_2(0)$.
Now Eq. \ref{1dSch} has a conserved Wronskian
\begin{equation} W = \chi_1(r) \chi_2'(r) - \chi_2(r) \chi_1'(r).
\end{equation}
It is easy to verify directly from the differential equation that
$W^\prime(r) = 0$; this is true (Abel's theorem) because there are no
$\chi^\prime$ terms in the differential equation. But comparing the
values of the conserved Wronskian at zero and $\infty$,
\begin{equation} W(\infty) = -k (\sin^2(k r + \delta) + \cos^2(k r + \delta)) = -k = W(0) = -A \chi_1'(0). \end{equation}
So then $|\chi_1'(0)| = k/|A|$, and our new expression for the
Sommerfeld enhancement, Eq. \ref{newenhance}, is just $S_k = 1/|A|^2$.
Now, our second form $S_k = |\chi(\infty)|^2$, where $\chi(r)$
satisfies the boundary conditions $\chi'(r) \rightarrow i k \chi(r)$
as $r \rightarrow \infty$, and $\chi(0) = 1$. By the asymptotic
behavior at large $r$, we can identify $\chi(r)$ as the linear
combination $\chi(r) = C (\chi_2(r) + i \chi_1(r))$, where $C$ is
some complex constant. But then $\chi(0) = C A = 1$, and $S_k =
|C|^2$, so as previously we obtain $S_k = 1/|A|^2$. Thus the two
formulae for the Sommerfeld enhancement are equivalent.

\subsection{Attractive  Coulomb Potential}

Let us see how this works in some simple examples. We are solving the
Schr\"{o}dinger equation for a particle of mass $M$ and asymptotic
velocity $v$, with potential
\begin{equation}
V(r) = - \frac{\alpha}{2r}
\end{equation}
which we solve with the boundary conditions of
Eqs. \ref{1dasymp1},\ref{1dasymp2}.

We simplify the analysis by working with the natural dimensionless
variables, with the unit of length normalized to the
Bohr radius, i.e. we take
\begin{equation}
r = \alpha^{-1} M^{-1} x
\end{equation}
Then the Schr\"{o}dinger equation becomes
\begin{equation}
- \chi^{\prime \prime} - \frac{1}{x} \chi = \epsilon^2 \chi
\end{equation}
where we have defined the parameter
\begin{equation}
\epsilon_v \equiv \frac{v}{\alpha}
\end{equation}

Of course we can solve this exactly in terms of hypergeometric
functions to find the result obtained by Sommerfeld
\begin{equation} S_k = \left|\frac{\frac{\pi}{\epsilon_v}}{1 -
e^{- \frac{\pi}{\epsilon_v}}}\right|
\end{equation}
Note that as $\epsilon_v \to \infty$, $S_k \to 1$ as expected; there
is no enhancement at large velocity. For the attractive Yukawa at
small velocities we have the enhancement
\begin{equation}
S_k \to \frac{\pi \alpha}{v}
\end{equation}
while for the repulsive case, there is instead the expected
exponential suppression from the need to tunnel through the Coulomb
barrier
\begin{equation}
S_k \sim e^{-\frac{\pi |\alpha|}{v}}
\end{equation}
To get some simple insight into what is going on, let us re-derive
these results approximately.  For $x$ much
smaller than $1/\epsilon_v^2$, we can ignore the kinetic term. In the WKB approximation, the waves are of the form $x^{1/4} e^{i \sqrt{x}}$, so the amplitudes grow like $x^{1/4}$. In
order to match to a unit norm wave near $x \sim 1/\epsilon_v^2$, we
have to scale the wavefunction at small $x$ by a factor $\sim
\epsilon_v^{1/2}$, so that near the origin
\begin{equation}
\chi \sim x \epsilon_v^{1/2} \sim \alpha M r \epsilon_v^{1/2}
\end{equation}
from which we can read off the derivative at the origin $\frac{d
\chi}{dr}(0) = \epsilon_v^{1/2} \alpha M $, and with $k = Mv$ we can
determine
\begin{equation}
S_k \sim |\frac{\epsilon_v^{1/2} \alpha M}{Mv}|^2 = \frac{\alpha}{v}
\end{equation}
which is correct parametrically. We can also arrive at this result
from the second form for $S_k$. The computation is even more direct
here. We have waveforms growing as $x^{1/4}$ towards $x \sim
\frac{1}{\epsilon_v^2}$. In the region near $x \sim 1/\epsilon_v$,
we must transition to a purely outgoing wave. This is a
transmission/reflection problem, and ingoing and outgoing waves from the
left will have comparable amplitude. When we continue these back to
the origin, we will then have an amplitude reduced by a factor $\sim
\epsilon_v^{1/2}$. Then,
\begin{equation}
S_k \sim \left(\frac{1}{\epsilon_v^{1/2}}\right)^2 \sim
\frac{\alpha}{v}
\end{equation}

\subsection{Attractive Well and Resonance Scattering}
Let us do another example, where
\begin{equation}
V = - V_0 \theta(L - r), V_0 \equiv \frac{\kappa^2}{2M}
\end{equation}
The solution inside is $\chi(r) = A \sin k_{in} r$, where $k_{in}^2 =
\kappa^2 + k^2$, while outside we write it as $\sin(k(r - L) +
\delta$. Then, matching across the boundary at $r=L$ gives
\begin{equation} A \sin k_{in} L = \sin \delta, \, k_{in} A \cos k_{in}
L = k \cos \delta
\end{equation} Squaring these equations and adding them we can
determine
\begin{equation}
A^2 = \frac{1}{\sin^2 k_{in} L + \frac{k^2_{in}}{k^2} \cos^2 k_{in} L}
\end{equation}
and so
\begin{equation}
S_k = \frac{A^2 k_{in}^2}{k^2} = \frac{1}{\frac{k^2}{k_{in}^2} \sin^2
k_{in} L + \cos^2 k_{in}L}
\end{equation}
Now for $k^2/2M \ll V_0$, we have $k_{in} = \kappa + k^2/(2 \kappa)
+ \cdots$. Clearly, if $\cos \kappa L$ is not close to zero, there is
no enhancement. However, if $\cos \kappa L = 0$, then we have a large
enhancement
\begin{equation}
S_k \to \frac{\kappa^2}{k^2}
\end{equation}
This has a very simple physical interpretation in terms of resonance
with a zero-energy bound state. Our well has a number of bound
states, and typically the binding energies are of order $V_0$. We
see that if $\cos \kappa L$ is not close to zero, we have to have $A
\sim k/k_{in}$ be small. However, if accidentally $\cos \kappa L =
0$, then there is a zero-energy bound state: the wavefunction can
match on to $\psi = 1$ for $r > L$ smoothly, giving a zero-energy
bound state. The enhancement is of the form
\begin{equation}
S \sim \frac{V_0}{E - E_{\text{bound}}} \sim \frac{\kappa^2}{k^2}
\end{equation}
Note this formally diverges as $v \to 0$, but is actually cut off by the finite width of the state as familiar from Breit-Wigner.

\subsection{Attractive Yukawa Potential}
Now let us examine
\begin{equation}
V(r) = -\frac{\alpha}{2r} e^{-m_\phi r}.
\end{equation} Working again
in Bohr units, we have \begin{equation} V(x) = - \frac{1}{x}
e^{-\epsilon_\phi x} \end{equation} where
\begin{equation}
\epsilon_\phi \equiv \frac{m_{\phi}}{\alpha M}.
\end{equation} Now,
if $\epsilon_\phi \ll \epsilon_v^2$, the Yukawa term is always
irrelevant and we revert to our previous Coulomb analysis.

However, if $\epsilon_\phi \gg \epsilon_v^2$, our analysis changes;
we will use the second expression for the Sommerfeld enhancement for
simplicity. The potential turns off exponentially around $x \sim
1/\epsilon_\phi$. Now, the effective momentum is
\begin{equation}
k^2_{\rm{eff}} = \frac{1}{x} e^{-\epsilon_\phi x} + \epsilon_v^2
\end{equation}
and the quantity
\begin{equation}
\left|\frac{k_{\text{eff}}^\prime}{k_{\text{eff}}^2} \right|
\end{equation}
determines the length scale the potential is varying over relative
to the wavelength; so long as it is small, the WKB approximation
is good, and we have a waveform growing as $k^{-1/2}_{\text{eff}} e^{i
\int^x  dx^\prime k_{\text{eff}}(x^\prime)}$. Note that for $1 \ll x \ll
1/\epsilon_\phi$, the WKB approximation is manifestly good. Let us
now take the arbitrarily low velocity limit, where $\epsilon_v \to
0$. Then in the neighborhood of $x \sim 1/\epsilon_\phi$ we have
$k^2_{\text{eff}} \sim \epsilon_\phi e^{-\epsilon_\phi x}$, and
\begin{equation}
\left| \frac{k_{\text{eff}}^\prime}{k_{\text{eff}}^2} \right| \sim
\sqrt{\epsilon_\phi} e^{\frac{1}{2} \epsilon_\phi x} \sim \frac{\epsilon_\phi}{k_{\text{eff}}}
\end{equation}
so the WKB approximation breaks down when $k_{\text{eff}} \sim
\epsilon_\phi$, where the WKB amplitude is $\sim
\epsilon_\phi^{-1/2}$. The potential then varies more sharply than
the wavelength, and we have a reflection/transmission problem, with
an O(1) fraction of the amplitude escaping to infinity. The
enhancement is then
\begin{equation}
S \sim \frac{1}{\epsilon_\phi} \sim \frac{\alpha M}{m_\phi}
\end{equation}
We did this analysis for $\epsilon_v \to 0$, but clearly it will
hold for larger $\epsilon_v$, till $\epsilon_v \sim \epsilon_\phi$,
at which point it matches smoothly to the $\frac{1}{\epsilon_v}$
enhancement we get for the Coulomb problem. The crossover with
$\epsilon_v \sim \epsilon_\phi$ is equivalent to $M v \sim m_\phi$,
when the deBroglie wavelength of the particle is comparable to the
range of the interaction. This is intuitive--as the particle
velocity drops and the deBroglie wavelength becomes larger than the
range of the attractive force, the enhancement saturates. Of course
if $\epsilon_\phi$ is close to the values that make the Yukawa
potential have zero-energy bound states, then the enhancement is
much larger; we can get an additional enhancement $\sim
\epsilon_\phi/\epsilon_v^2$ up to the point where it gets cut off by
finite width effects.

\begin{figure}[ht]
\centering
\includegraphics[width=0.5\textwidth]{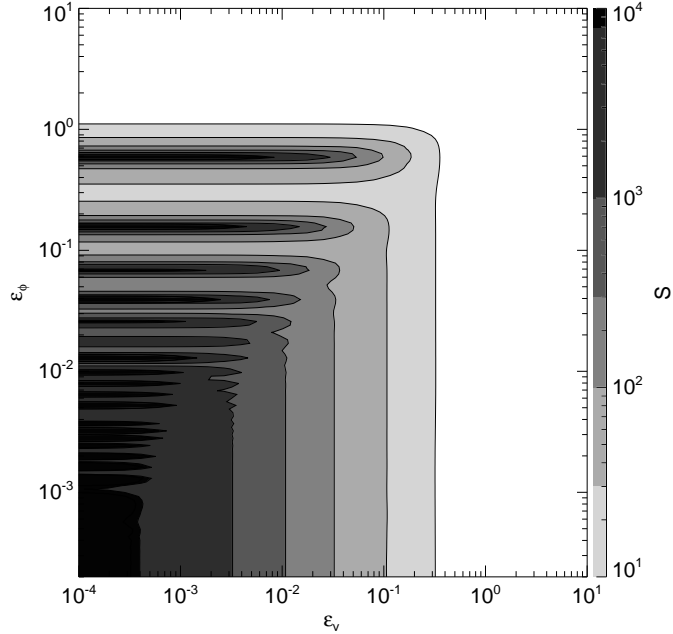}
\caption{Contour plot of $S$ as a function of $\epsilon_\phi$ and $\epsilon_v$. The lower right triangle corresponds to the zero-mass limit, whereas the upper left triangle is the resonance region.}
\label{contourplot}
\end{figure}

\begin{figure}[ht]
\centering
\includegraphics[width=0.5\textwidth]{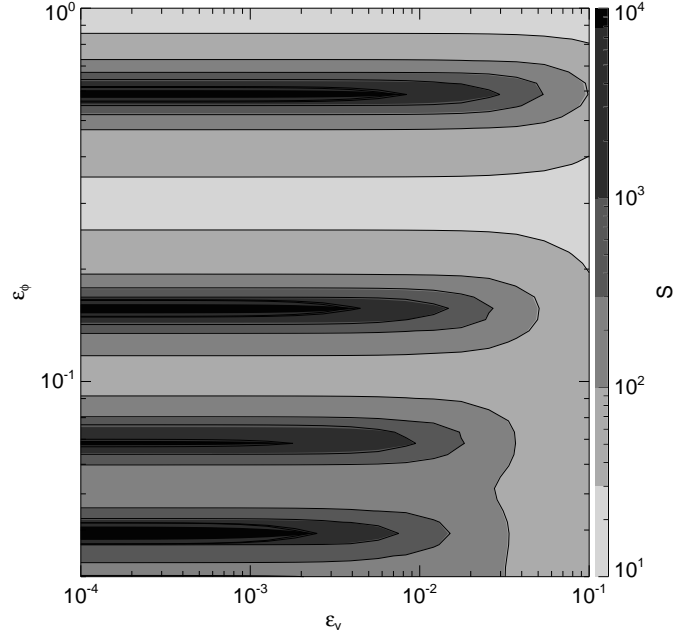}
\caption{As in Fig. \ref{contourplot}, showing the resonance region in more detail.}
\label{contourplot_resonance}
\end{figure}

In this simple theory it is of course also straightforward to solve
for the Sommerfeld enhancement numerically.  We show the enhancement
as a function of $\epsilon_\phi$ and $\epsilon_v$ in Figs.
\ref{contourplot} and \ref{contourplot_resonance}.

\subsection{Two-particle annihilation}
Let us finally consider our real case of interest, involving
two-particle annihilation. To keep things simple, let us imagine that
the two particles are not identical, for instance they could be
Majorana fermions with opposite spins; we can restrict to this case
because none of the interactions we consider depend on spin, and the
annihilation channels we imagine can all proceed from this spin
configuration. Let us imagine that there are a number of states
$\chi_A$ of (nearly) equal mass, with the label $A$ running from $A
= 1, \cdots, N$. The two-particle Hilbert space is labeled by states
$|\vec{x},A;\vec{y},B\rangle$ and a general wavefunction is $\langle
\psi|\vec{x},A;\vec{y},B\rangle \equiv \psi_{AB}(\vec{x},\vec{y})$.
There is some short-range annihilation Hamiltonian $U
\delta^3(\vec{x}_1 - \vec{x}_2)$ into decay products $|final
\rangle$; where $U$ is an operator
\begin{equation}
\langle final | U |A B \rangle \equiv M^{ann}_{AB}
\end{equation}
Now, suppose there is also a long-range interaction between the two
particles with a potential. The general Schr\"{o}dinger equation is of
the form
\begin{equation}
-\frac{1}{2M} \left(\nabla_x^2 + \nabla_y^2\right) \psi_{AB}(x,y) +
V_{ABCD}(x - y) \psi_{CD}(x,y) = E \psi_{AB}(x,y)
\end{equation}
As usual we factor out the center-of-mass motion by writing
$\psi_{AB}(\vec{x},\vec{y}) = e^{i \vec{P} \cdot (\vec{x} +
\vec{y})} \phi_{AB}(\vec{x} - \vec{y})$, and we have
\begin{equation}
-\frac{1}{M} \nabla_r^2 \phi_{AB}(\vec{r}) + V_{ABCD}(\vec{r})
\phi_{CD}(\vec{r}) = E_{CM} \phi_{AB}(\vec{r})
\end{equation}
where $E_{CM}$ is the energy in the center-of-mass frame.

We have an initial state with some definite $A_*,B_*$, and we take
as an unperturbed solution
\begin{equation}
\phi^{(A^* B^*) (0)}{AB} = \delta^{A_*}_A \delta^{B^*}_B e^{i k z}
\end{equation}
The annihilation cross section without the interaction $V$ is
proportional to
\begin{equation}
\sigma^{(0)}_{ann} \propto |M^{ann}_{A^* B^*}|^2
\end{equation}
However with the new interaction, the annihilation cross section is
\begin{equation}
\sigma_{ann} \propto |\phi^{(A^* B^*)}_{CD}(0) M_{CD}|^2
\end{equation}
So we can write
\begin{equation}
\sigma_{ann} = \sigma_{ann}^{(0)} \times S_k^{A^* B^*}
\end{equation}
where
\begin{equation}
S_k^{A^* B^*} = \frac{|\phi^{A^* B^*}_{CD}(0)
M^{ann}_{CD}|^2}{|\phi^{A^* B^*}_{CD}(\infty) M^{ann}_{CD}|^2}
\end{equation}
Just as above, in reducing the problem to an $S$-wave computation, we can replace $\phi^{A^* B^*}_{CD}(0)$ with $(\chi^{\prime}(r=0))^{A^* B^*}_{CD}$ in the obvious way.

Of course one contribution to $V_{ABCD}$ comes from the small mass
splittings between the states. If we write the (almost equal) common
mass term for the DM states as $(M + \Delta M_A)  \chi_A \chi_A$,
the $\Delta M$'s show up  in the potential as
\begin{equation}
V^{split}_{ABCD} = (\Delta M_A + \Delta M_B) \delta_{AC} \delta_{BD}
\end{equation}
For the Sommerfeld enhancement, we also need some
long-range attractive interaction. As we have discussed, vectors are possibly the most promising candidate.
The leading coupling to spin-1 particles $a_{\mu i}$ is
\begin{equation}
g \bar{\chi}_{A} \bar \sigma^\mu \chi_B T^{i}_{AB} a_{i \mu}
\end{equation}
and ignoring the mass of the gauge boson this gives us a $1/r$ contribution to the effective potential
\begin{equation}
V^{gauge}_{ABCD}(\vec{r}) = - \alpha \frac{1}{r} T^{i}_{AC}
T^{i}_{BD}
\end{equation}
while taking the vector masses into account gives both Yukawa
exponential factors and a more complicated tensor structure. In
total,
\begin{equation}
V_{ABCD} = V^{split}_{ABCD} + V^{gauge}_{ABCD}
\end{equation}
Now, it is obvious that in our basis, the $T^{i}_{AB}$ should be antisymmetric
 \begin{equation} T^{i}_{AB} = -
T^{i}_{BA}
\end{equation}
since the gauge symmetry must be a subgroup of the $SO(N)$ global symmetry preserved by the large common mass term $M \chi_A \chi_A$, and the $SO(N)$ generators are antisymmetric. Thus, the coupling to vectors in this basis is necessarily
off-diagonal.
Let us look at a simple example, where $N=2$ and we have a single Abelian gauge field.
The $\phi_{AB}$ span a 4-dimensional Hilbert space, with
$|11 \rangle,|12\rangle,|21\rangle,|22\rangle$ as a basis.
Since the gauge boson exchange necessarily changes $1 \leftrightarrow 2$, $V_{ABCD}$ is
 block diagonal, operating in two separate Hilbert spaces, spanned by $(|11\rangle, |22\rangle)$ and $(|12\rangle,|21\rangle)$.
 Since we are ultimately interested in scattering with $11$ initial states, let us look at the first one, where we have
\begin{equation}
{\bf V} = \left(\begin{array}{cc} 2 \Delta M & -\frac{\alpha}{r} \\ -\frac{\alpha}{r} & 0 \end{array} \right), \quad \text{in the basis} \quad |11 \rangle =  \left( \begin{array}{c} 0 \\ 1 \end{array} \right),  |22 \rangle = \left( \begin{array}{c} 1 \\ 0 \end{array} \right). 
\end{equation}
Clearly, if $\Delta M$ is enormous, we will not have any interesting Sommerfeld enhancement in (11) scattering, since in this case there is no long-range interaction between 11 at all, so let us assume that $\Delta M$ is smaller than the kinetic energy of the collision.
Now it is clear that as $r \to \infty$, the mass splitting dominates the potential, and obviously particle 1 is the lightest state. However, at smaller distances, the gauge exchange term dominates. This is not diagonal in the same basis, and has ``attractive'' and ``repulsive'' eigenstates with energies $\pm \alpha/r$.  Note that the asymptotic $|11 \rangle$ state is an equal linear combination of the attractive and repulsive channels. While the repulsive channels suffer a Sommerfeld suppression, the attractive channel is Sommerfeld enhanced. Note that as long as $\Delta M$ is parametrically smaller than the kinetic energy, its only role in this discussion was to split the two asymptotic states, and therefore determine what the natural initial states are. Note also that if $\Delta M$ is large but not infinite, the mixing with $2$ generates an attractive potential between $11$ of the form $V_{eff}(r) = - \alpha^2/(\Delta M r)$. It would be interesting to understand  these limits of the multistate Sommerfeld effect in parametric detail; we defer this to future work.

It is very easy to see that our conclusion about the presence of a
Sommerfeld effect is general for any gauge interaction. We think of
$V^{gauge}_{(AB)(CD)}$ as a matrix in the Hilbert space spanned by
$(AB)$. Note that since the $T^i$ are antisymmetric, they are also
traceless, and as a consequence, the matrix $V^{gauge}_{(AB)(CD)}$
is also traceless and so has both positive and negative eigenvalues,
reflecting the obvious fact that gauge exchange gives us both
attractive and repulsive potentials. Let us go to a basis in $(AB)$
space where $V^{gauge}$ is diagonal, and denote eigenvectors with
negative (attractive) eigenvalues as $f^{attractive}_{AB}$ and
repulsive ones as $f^{repulsive}_{AB}$. We can think of the initial
wavefunction in $AB$ space $\delta^{A^* B^*}_{AB}$ as a state in
$(AB)$ space and expand it in terms of these eigenvectors as
\begin{equation}
\delta^{(A^* B^*)}_{AB} = C^{A^* B^*}_{att} f^{attractive}_{AB} +
C^{A^* B^*}_{rep} f^{repulsive}_{AB}
\end{equation}
we can determine the coefficients by dotting left-hand side and right-hand side into the
eigenvectors, so that
\begin{equation}
\delta^{(A^* B^*)}_{AB} = f^{attractive}_{A^* B^*}
f^{attractive}_{AB} + f^{repulsive}_{A^* B^*} f^{repulsive}_{AB}
\end{equation}
Then, since the repulsive components are exponentially suppressed at the
origin while the attractive components are enhanced, we get a Sommerfeld
enhancement as long as $f_{A^* B^*} \neq 0$. In particular, for
scattering the same species, this is true so long as $f_{A^* A^*}
\neq 0$. Said more colloquially, these Majorana states are linear
combinations of ``positive'' and ``negative'' charged states; so long
as they have any component which would mutually attract, there is a
Sommerfeld enhancement from that component alone.

\onecolumngrid
\bibliography{sommerfeld}
\bibliographystyle{apsrev}

\end{document}